# Coherent Sub-Nanosecond Switching of Perpendicular Magnetization by the Field-like Spin-Orbit Torque without an External Magnetic Field

William Legrand, Rajagopalan Ramaswamy, Rahul Mishra, and Hyunsoo Yang[*]

*Department of Electrical and Computer Engineering, National University of Singapore, 117576, Singapore*

We theoretically study the influence of a predominant field-like spin-orbit torque on the magnetization switching of small devices with a uniform magnetization. We show that for a certain range of ratios (0.23-0.55) of the Slonczewski to the field-like torques, it is possible to deterministically switch the magnetization without requiring any external assist field. A precise control of the pulse length is not necessary, but the pulse edge sharpness is critical. The proposed switching scheme is numerically verified to be effective in devices by micromagnetic simulations. Switching without any external assist field is of great interest for the application of spin-orbit torques to magnetic memories.

[*]eleyang@nus.edu.sg

# I. INTRODUCTION

The recent discovery of domain wall propagation [1-3], current-induced switching [4-8] and high-frequency magnetization oscillations [9-11] induced by spin-orbit torque (SOT) in thin-film nanostructures has led to considerable research interest in the field of spintronics. It is notably expected that SOT allows high-speed, low-energy switching of magnetic devices which could benefit many technological applications, especially magnetic random access memories (MRAM). SOT switching of nanomagnets with either in-plane [6,8] or out-of-plane [5,7,12] magnetization has been investigated mostly in trilayered HM/FM/Ox structures where a heavy metal (HM) is adjacent to a ferromagnetic (FM) layer capped by an oxide (Ox). In this configuration, an in-plane current can be applied to the trilayer to generate a perpendicular spin-density from spin-orbit interaction, which is transferred to the FM and leads to switching. This perpendicular spin-density generation is currently attributed either to the spin Hall effect [2,7,13] in the HM layer or to the Rashba effect [1,5,14-17] at the FM/Ox interface, or both, and is still under debate. Regardless of the origin of the torque, the injected spin polarization is expected to be in-plane, so that compared to spin-transfer torque in the previous configuration with perpendicular anisotropy [18], SOT allows faster, non-precessional switching due to maximum initial torque-efficiency.

However, as the injected spin-polarization simply drives the magnetization in the plane, an external in-plane assist field has to be applied parallel to the current in order to achieve deterministic switching. The direction of switching is defined by the relative directions of both external field and applied current [5,7,19]. A recent study [20] has also underlined the essential role of the assist field to overcome the Dzyaloshinskii-Moriya interaction (DMI) in large devices during the non-uniform switching process. It is possible to obtain an external assist field in devices by engineering the magnetic stack, but this approach is hindered by disadvantages in terms of fabrication difficulties, efficiency, and stability. In fact, SOT switching without external field has been recently demonstrated in specific structures with broken lateral inversion symmetry [21] or by engineering a tilted anisotropy [22]. Nevertheless, the ability to perform SOT switching without assist field in standard magnetic stacks is of great interest for applications.

Current-induced SOT is often explained in terms of the field-like and the Slonczewski torques [1,7,23]. Recent experimental studies [24-28] found that the field-like torque can be much larger than the Slonczewski torque in some material combinations, and this field-like torque is also capable of switching the magnetization [29]. Here, we investigate the potential of SOT switching in the case where the field-like torque is predominant and combined with the Slonczewski torque. This study is focused on dynamic switching in small single-domain perpendicularly magnetized nanomagnets (typically, $50\times50$ nm$^2$ or less), which are of significant interest for applications. Our results show that for certain relative values of the Slonczewski and field-like torques, a deterministic switching of the magnetization can be achieved without assist field. This switching is not sensitive to the pulse length but rather to the pulse rising time. Micromagnetic simulations have then been performed to assess the suitability of this new approach for real devices, and to argue in favor of experimental feasibility.

SOT can be described [23] as the sum of the longitudinal (Slonczewski) torque $\boldsymbol{\tau}^{\parallel} \sim \mathbf{m}\times(\hat{\boldsymbol{\sigma}}\times\mathbf{m})$ and the transverse (field-like) torque $\boldsymbol{\tau}^{\perp} \sim \hat{\boldsymbol{\sigma}}\times\mathbf{m}$, where $\mathbf{m}$ denotes the direction of the magnetization vector and $\hat{\boldsymbol{\sigma}}$ collinear to $\hat{\mathbf{y}}$ is the in-plane polarization of the spin current, perpendicular to the current-density $\mathbf{J} = J_e \hat{\mathbf{x}}$ as shown in Fig. 1(a). Thus the evolution of the magnetization can be described by the Landau-Lifshitz-Gilbert equation, modified with the additional SOT contributions

$$\frac{\partial \mathbf{m}}{\partial t} = -\gamma \mu_0 \mathbf{m} \times \mathbf{H}_{\text{eff}} + \alpha \mathbf{m} \times \frac{\partial \mathbf{m}}{\partial t} + \boldsymbol{\tau}^{\|} + \boldsymbol{\tau}^{\perp} \quad (1)$$

where $\mathbf{m}$ is described in the polar coordinates $(\theta, \varphi)$, with the gyromagnetic ratio $\gamma$ ($1.76 \times 10^7$ Oe$^{-1}$ s$^{-1}$), the Gilbert damping constant $\alpha$, and the effective field $\mathbf{H}_{\text{eff}} = H_{ani} \cos\theta \hat{\mathbf{z}} + \mathbf{H}_{\text{ext}}$ composed of both effective anisotropy field and any external field applied along $\hat{\mathbf{x}}$, $\mathbf{H}_{\text{ext}} = H_x \hat{\mathbf{x}}$. We describe $\boldsymbol{\tau}^{\perp}$ and $\boldsymbol{\tau}^{\|}$ by their efficiencies $c^{\perp}$ and $c^{\|}$,

$$\boldsymbol{\tau}^{\perp} = \frac{\gamma \hbar c^{\perp}}{2|e|M_S t_{FM}} J_e \left( \hat{\boldsymbol{\sigma}} \times \mathbf{m} \right) \quad (2)$$

and

$$\boldsymbol{\tau}^{\|} = \frac{\gamma \hbar c^{\|}}{2|e|M_S t_{FM}} J_e \left[ \mathbf{m} \times \left( \hat{\boldsymbol{\sigma}} \times \mathbf{m} \right) \right] \quad (3)$$

where $M_S$ and $t_{FM}$ are the saturation magnetization and the thickness of the FM layer, respectively. Even if both these torques do not necessarily arise from the spin Hall effect, we use this notation for comparison of efficiencies with spin Hall angles reported in literature. Within this description, a positive value of $c^{\|}$ corresponds to a positive, equal value of the spin Hall angle [7]. It is also important to note that the Slonczewski torque does not necessarily compete with damping, and can contribute or oppose to damping proportionally to the cosine of the angle between $\hat{\boldsymbol{\sigma}}$ and $\mathbf{m}$, denoted by $\cos(\hat{\boldsymbol{\sigma}}, \mathbf{m})$ [30].

We can rescale the external field and the torques by the anisotropy effective field $H_{ani}$ so that Eq. (1) is given in a convenient dimensionless form

$$\frac{\partial \mathbf{m}}{\partial t'} = -\mathbf{m} \times (\cos(\theta)\hat{\mathbf{z}} + h_x\hat{\mathbf{x}}) + \alpha \mathbf{m} \times \frac{\partial \mathbf{m}}{\partial t'} - h^{\parallel}\left[\mathbf{m} \times (\mathbf{m} \times \hat{\boldsymbol{\sigma}})\right] - h^{\perp}(\mathbf{m} \times \hat{\boldsymbol{\sigma}}) \quad (4)$$

where $t' = \gamma\mu_0 H_{ani} t$ is a rescaled dimensionless time, $h_x = H_x / H_{ani}$, and $h^{\parallel,\perp} = H^{\parallel,\perp} / H_{ani} = \left(\hbar J_e c^{\parallel,\perp}\right)/\left(2|e|\mu_0 M_S t_{FM} H_{ani}\right)$ are the spin-orbit torque effective fields.

As can be seen from Eq. (4), $\boldsymbol{\tau}^{\perp}$ generates a precession of **m** around its equivalent field $h^{\perp}\hat{\boldsymbol{\sigma}}$, and can then potentially lead to switching if the magnetization vector crosses the *xy*-plane. Previous theoretical works [31,32] modelled the current-induced SOT switching considering $\boldsymbol{\tau}^{\parallel}$ only, thus focusing on direct switching without precession. Other recent works [33] studied coherent switching in small nanomagnets with an in-plane assist field and a moderate $\boldsymbol{\tau}^{\perp}$ only. They underlined the oscillatory switching behavior depending on the length and intensity of the current pulse, similar to the behavior of orthogonal spin-torque (OST) devices [34,35]. To be able to deterministically switch the ferromagnet in this scheme, the current pulse needs to be precisely ended once the magnetization has reversed. On the contrary, a precise control of the pulse length is not necessary in our scheme, as discussed below.

To study the role of a dominant field-like torque, we solve Eq. (4) assuming a uniform magnetization in a small nanomagnet (for example with dimensions of $w \times w = 50 \times 50 \text{ nm}^2$) and first discuss the trajectories qualitatively. To get a closer look into the expected behavior for real devices, we also give some illustrative numerical results using the magnetic parameters of CoFeB such as $M_S = 1500 \text{ emu cm}^{-3}$, $\alpha = 0.01$, and $t_{FM} = 1.6$ nm. $H_{ani}$ is set to 1000 Oe unless otherwise stated, which allows a stability factor up to $\Delta = \mu_0 M_S H_{ani} w^2 t_{FM}/2k_B T \sim 70$ at room temperature. We neglect all thermal contributions during switching. With these parameters, one can express $h^{\parallel,\perp} = c^{\parallel,\perp}(J_e/J_0)$ with a characteristic current density $J_0 = (2|e|\mu_0 M_S t_{FM} H_{ani})/\hbar = 7.28 \times 10^7$ A cm$^{-2}$.

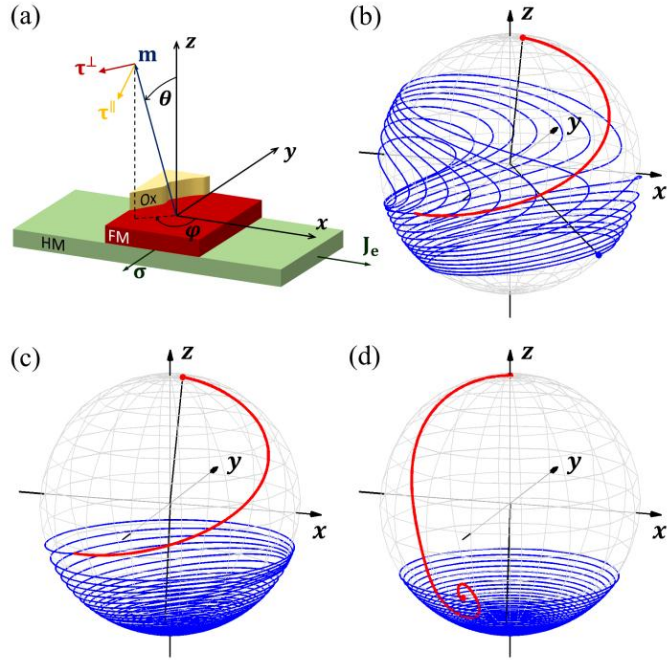

FIG. 1. (a) A sectional view of the trilayered system and directions of the different torques. (b)-(c) Field-assisted trajectory of the magnetization with $t'_{rise} = 40\pi$, $h_x = 0.1$, $h^{\parallel} = 0.45$ and $h^{\perp} = 0$ (b), $h^{\perp} = 0.005$ (c). (d) Switching trajectory without assist field ($h_x = 0$) with $t'_{rise} = 0$, $h^{\parallel} = 0.2$, and $h^{\perp} = 0.8$. The red (blue) curve represents the dynamics of **m** under a current pulse (for the subsequent relaxation), and the red (blue) dot indicates the original direction of **m** (direction after applying the pulse and relaxation). Only relaxation within 20 ns is shown for clarity.

## II. QUASI-STATIC REGIME

We first briefly recall the switching scenario under assist field, when $\boldsymbol{\tau}^{\perp}$ is not taken into account or is negligible [31,36]. The anisotropy allows two equilibrium positions for the magnetization, a 'up' (**m** pointing close to $+\hat{\mathbf{z}}$) and a 'down' (**m** pointing close to $-\hat{\mathbf{z}}$) equilibria, both slightly tilted away from the *z*-axis because of the in-plane external assist field. Let us consider that the initial position of **m** is at the 'up' equilibrium position. As the current gradually increases, the 'up' and 'down' stable equilibria are progressively driven closer to $+\hat{\mathbf{x}}$ and $-\hat{\mathbf{x}}$, respectively, while staying in the *xz*-plane. When the critical (positive) value of the current is reached, the 'up' position is suppressed and **m** switches to an 'intermediate' equilibrium away from the *xz*-plane and close to $\hat{\boldsymbol{\sigma}}$ [31]. The assist field is actually not required to suppress the 'up' equilibrium. Without external field, the critical value of the torque to suppress both 'up' and 'down' equilibria is identical, $h_{u,d}^{\parallel} = 0.5$ [36]. The role of the assist field is to induce asymmetry between the ranges of $h^{\parallel}$ for which the 'up' and 'down' equilibria exist (with $h_d^{\parallel} > 0.5$ and $h_u^{\parallel} < 0.5$), so that the 'down' position is still a stable equilibrium when the 'up' state is no longer an equilibrium. After the pulse falling edge, **m** eventually relaxes to the 'down' position.

To achieve the opposite switching event (i.e., from 'down' to 'up') through a symmetric process, either an opposite value of the assist field or an opposite value of the current needs to be applied. This scenario relies on the quasi-static evolution of **m** and then describes accurately the most-studied case of SOT switching using either a DC-current or a current-pulse of rising time much longer than the characteristic time of precession ($t' = 2\pi$, that is corresponding to $t \approx 0.36$ ns in our case). In this work, the current is modeled as a trapezoidal pulse with equal rising and falling time, $t_{rise}$. When the corresponding value $t'_{rise} = \gamma\mu_0 H_{ani} t_{rise}$ is much greater than $2\pi$, switching occurs as described above following quasi-static evolution.

The scenario of switching due to a slowly varying current pulse is exemplified in Fig. 1(b), which shows the switching trajectory for $h^{\parallel} = 0.45$, $h_x = 0.1$, and $t'_{rise} = 40\pi \gg 2\pi$. The switching trajectory to the intermediate equilibrium is represented by the red curve, and the subsequent relaxation by the blue curve. Assuming a macrospin behavior in the case of a Slonczewski torque only, the relaxation event disturbs the switching process. As revealed by numerical simulations [31] and as can be understood from the many precessions during the relaxation in Fig. 1(b), the final position between 'up' and 'down' is extremely sensitive to the values of $\alpha$ and $h^{\parallel}$.

Such a non-predictable relaxation is due to the hysteretic behavior of the transition to the 'intermediate' position [36], that involves dynamics during the relaxation even in the case of a slow evolution of the applied current. With the parameters of Fig. 1(b), the minimum value of $h^{\parallel}$ for the intermediate position to be stable in our simulations, which we define as $h_i^{\parallel} \approx 0.325$, is significantly lower than $h_u^{\parallel} \approx 0.43$ (one can derive the same values from Ref. [36]). After switching to the 'intermediate' position when $h^{\parallel} > h_u^{\parallel}$, **m** is destabilized only when $h^{\parallel} < h_i^{\parallel}$, which happens during the falling edge of the pulse. At this time, both 'up' and 'down' equilibria are stable again and located far from $\hat{\boldsymbol{\sigma}}$, so that **m** can relax to either one through a large dynamic precession covering both $z < 0$ and $z > 0$ half-spheres [Fig. 1(b)]. The unpredictability of the dynamics to the final position is amplified even more by the low value of $\alpha$, as the precession amplitude remains large after the current falling edge so that many oscillations occur, resulting in non-deterministic switching.

Considering the above hysteretic behavior of the transition to the 'intermediate' position brings two conclusions. (i) It explains clearly why a higher $\alpha$ [31,33] that damps the oscillations during relaxation from the 'intermediate' state (and also destabilizes the intermediate equilibrium earlier) enables the deterministic field-assisted switching. By increasing $\alpha$ to 0.04 (not shown) in the situation of Fig. 1(b), even though switching occurs through the 'intermediate' equilibrium, the relaxation trajectory is damped enough to remain in the $z<0$ half-sphere and switching is deterministic. (ii) As the 'intermediate' position is shifted and modified by the in-plane assist field, $h_x$ also contributes to the destabilization of the 'intermediate' equilibrium, and one can refer to Ref. [36] to find the minimum value of $\alpha$ so that $h_u^{\|} < h_i^{\|}$ at a given $h_x$. When $\alpha$ is greater than this minimum, the 'intermediate position' is unstable when the 'up' equilibrium is suppressed, and **m** directly switches to the 'down' equilibrium position ($h^{\|} < h_d^{\|}$), ensuring deterministic switching. For $h_x \approx 0.15$ or higher, the 'intermediate' equilibrium is unstable at $h_u^{\|}$ regardless of $\alpha$, and switching directly proceeds to the final $-\hat{\mathbf{z}}$ direction (not shown).

We extend to the influence of the field-like torque as well, and find that the transverse field from $\boldsymbol{\tau}^{\perp}$ allows deterministic quasi-static switching, even with a low $\alpha$. As the 'intermediate' position is shifted and modified by the field-like torque as well, it is destabilized much earlier and the relaxation proceeds only in the $z < 0$ half sphere. Keeping the parameters of Fig. 1(b) and $\alpha = 0.01$ but adding a small field-like torque ($h^{\perp} = 0.005$), switching is found to be deterministic, as shown in Fig. 1(c). These results indicate that it is necessary to consider the dynamics and the field-like term, as they may have a critical influence on the SOT switching.

## III. DYNAMIC REGIME

### A. Switching mechanism

The role of $\boldsymbol{\tau}^\perp$ can be much more significant than enhancing the field-assisted switching, as we show that a predominant field-like torque enables switching without any assist field. Figure 1(d) shows one example of switching without assist field ($h_x = 0$), in the case of a dominant field-like torque with $h^\parallel = 0.2$ and $h^\perp = 0.8$, corresponding to $c^\parallel / c^\perp = 0.25$. The rising time $t_{rise}$ is assumed as infinitely small and thus negligible as compared to the characteristic period of precession of the system, so that the dynamics of **m** dominates switching. As can be seen from the red part of the trajectory, **m** precesses around $\hat{\boldsymbol{\sigma}}$ to cross the *xy*-plane at $t' \approx 0.81\pi$ or $t \approx 145$ ps, which is about half of the characteristic precession time, and directly stabilizes at the 'down' equilibrium position (after $t' \approx 5.6\pi$ or $t \approx 1$ ns). The blue part of the trajectory corresponds to the final relaxation after the falling edge of the pulse and indicates that **m** has finally switched.

In order to understand the switching mechanism, first consider the case of a sole field-like torque. Due to the sharp increase of the current pulse, **m** is still parallel to $\hat{\mathbf{z}}$ when the maximum-level of the current pulse is reached, inducing a large lag between **m** and the 'up' equilibrium displaced by the spin-orbit torques. This lag actually raises the energy of the system relative to the new energy minima corresponding to the displaced 'up' and 'down' equilibrium positions, and any dissipation would need some time to occur. Moreover, because of the field-like torque effective field being a transverse field, the 'up' and 'down' equilibrium positions are tilted away from the *xz*-plane towards $\hat{\boldsymbol{\sigma}}$. For a sufficiently high value of the current, the trajectory of constant energy starting from $+\hat{\mathbf{z}}$ and winding around the displaced 'up' equilibrium merges with the symmetric orbit winding around the displaced down 'equilibrium'. This creates a large trajectory of precession which conjugates the two 'up' and 'down' states by winding around them [Fig. 2(a)]. The critical $h^\perp$ value ($h^\perp = 0.5$) to observe this merging corresponds to when the homoclinic orbits of the system that start from $\hat{\boldsymbol{\sigma}}$ reach $\pm\hat{\mathbf{z}}$, when $E(\pm\hat{\mathbf{z}}) = E(\hat{\boldsymbol{\sigma}})$ with the energy density of the system derived from the anisotropy and field-like torque effective fields, $E(\mathbf{m}) = -\mu_0 M_S H_{ani}\left[(\mathbf{m}\cdot\hat{\mathbf{z}})^2/2 + h^\perp(\mathbf{m}\cdot\hat{\boldsymbol{\sigma}})\right]$.

Such a precessional motion from the field-like spin-orbit torque scenario is similar to the well-known field-pulse induced switching [37]. The precession of **m** is progressively damped while the current pulse is maintained [Fig. 2(b)], until **m** finally falls into one hemisphere, leading to either switching or non-switching in a non-predictable manner. By adjusting the length of the pulse, the current can be stopped at the time when **m** reaches the 'down' equilibrium, ensuring that switching is achieved. However, such a switching scheme may not be practical for applications, because the switching pulse length needs to be modified according to any variations of the layer magnetic properties, field efficiency, and current. It was also found that due to the Gilbert damping of the precession, switching back can be prevented after switching from one equilibrium to another, so that a reliable, deterministic switching of the magnetization can be obtained for any pulse length longer than the required switching time [37]. However, as this regime is damping-dominated, the shorter switching times are not reachable. The well-balanced combination of the Slonczewski and field-like torques address these limitations and leads to the fast and deterministic switching observed in Fig. 1(d).

By choosing $c^\parallel/c^\perp$ significantly lower than unity but not negligible, the Slonczewski torque has a limited but crucial influence on switching, as we discuss below.

Different cases with increasing the Slonczewski torque are shown in Figs. 2(c)-2(f). As the field-like torque dominates, **m** precesses around $\hat{\boldsymbol{\sigma}}$ and the two equilibrium positions 'up' and 'down' are shifted towards $\hat{\boldsymbol{\sigma}}$, which is similar to the previous case of the sole field-like torque shown in Fig. 2(b). The 'up' and 'down' equilibrium positions under currents are also slightly shifted along the *x*-axis by the Slonczewski torque, as it is noticeable in Figs. 2(c)-2(f). However, the nature of the trajectories remains unchanged, and **m** still crosses the *xy*-plane at some point. Along the trajectory of **m** precessing around $\hat{\boldsymbol{\sigma}}$, which is located on the half sphere with $y < 0$ in all cases, the Slonczewski torque always generates a positive damping as $\cos(\hat{\boldsymbol{\sigma}}, \mathbf{m})$ has a positive sign. In analogy with the case of the field-pulse induced switching [37], it is natural to define a precession order *N* which is an half-integer counting the number of rotations of **m** around $\hat{\boldsymbol{\sigma}}$. By increasing the Slonczewski torque efficiency, the effective damping is consequently enhanced. As a result, the additional energy obtained from the lag is dissipated faster, limiting *N* to lower values for larger $c^{\parallel}$ [Figs. 2(b)-2(f)]. In the particular case of Fig. 2(f), the damping on the trajectory is so strong that **m** never crosses the *xy*-plane. Back to the example of Fig. 1(d) where $c^{\parallel} / c^{\perp} = 0.25$, we now understand that the Slonczewski torque is strong enough to efficiently damp the precession arising from the field-like torque, so that **m** is rapidly stabilized [38] at the 'down' equilibrium for the remaining part of the pulse, resulting in *N* = 1/2. As there is no assist field, the precession during relaxation cannot assist **m** to cross again the *xy*-plane backwards, and **m** eventually relaxes to $-\hat{\mathbf{z}}$ after the pulse. In this scheme, switching occurs without a requirement of an external assist field, within the sub-ns range, and through only a half-rotation (*N* = 1/2) around $\hat{\boldsymbol{\sigma}}$ because the Slonczewski torque prevents from switching back.

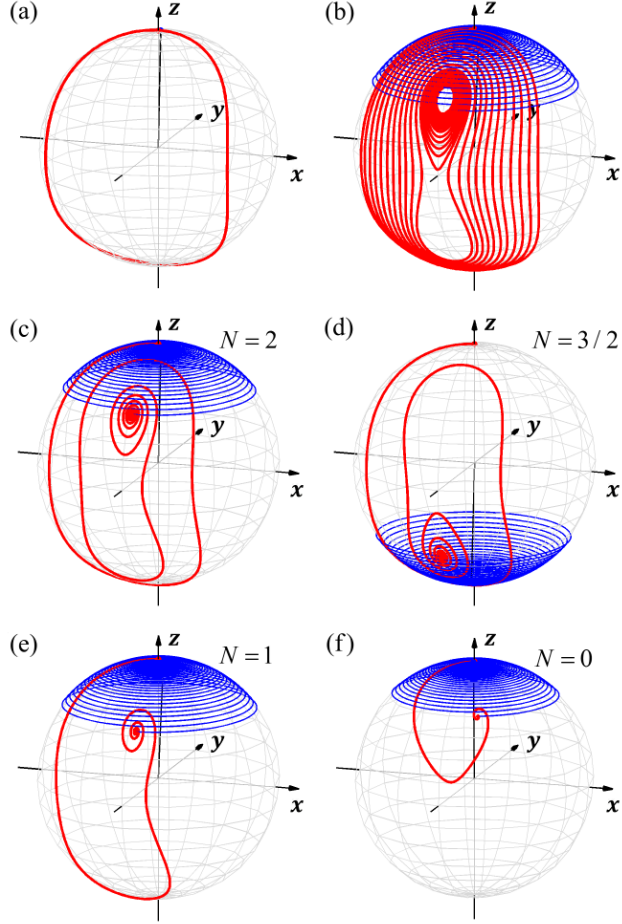

FIG. 2. (a) Trajectory obtained in the event of conservation of the energy with $\alpha = 0$ and $t_{rise} = 0$, that is, the isoenergetic orbit. Relaxation is not shown. (b) Dissipative trajectory obtained with $\alpha = 0.01$. Under currents, **m** finally falls into one hemisphere and relaxes to $+\hat{\mathbf{z}}$ in this case, but the final state 'up' or 'down' is not controllable. For both (a) and (b), $h^{\parallel} = 0$ and $h^{\perp} = 0.8$. (c)-(f) Trajectories for various values of $h^{\parallel}$ and $h^{\perp}$; (c) $h^{\parallel} = 0.04$ and $h^{\perp} = 0.8$, (d) $h^{\parallel} = 0.05$ and $h^{\perp} = 0.8$, (e) $h^{\parallel} = 0.1$ and $h^{\perp} = 0.8$, and (f) $h^{\parallel} = 0.3$ and $h^{\perp} = 0.6$. The precession order $N$ is indicated in each case. For (b)-(f), pulse length is 20 ns with no rising time, and relaxation within only 20 ns is shown. In each plot, the red (blue) curve represents the dynamics of **m** under the current pulse (for the subsequent relaxation).

In addition, this switching process is not sensitive to the length of the current pulse. As the magnetization directly stabilizes along the switched equilibrium direction under currents [see Fig. 1(d)], any pulse length longer than the switching time is able to deterministically switch the magnetic layer. Thus, contrary to the OST switching scheme [34,35] and field pulse induced magnetic switching [37,39], our proposed scheme does not require a precise control of the pulse length. Moreover, we have changed $\alpha$ from 0.001 to 0.2 in the simulations and found that this half-precession switching mode can be effective for any values of $\alpha$, because the necessary damping is obtained and adjusted from the Slonczewski torque. However, a high $\alpha$ is still beneficial, as it helps **m** to reach the $-\hat{\mathbf{z}}$ direction faster after the pulse falling edge. A last notable point is that in this scheme the two equilibrium directions are conserved at all times, and are equivalent for any current polarity. Both $\hat{\mathbf{z}}$ to $-\hat{\mathbf{z}}$ and $-\hat{\mathbf{z}}$ to $\hat{\mathbf{z}}$ toggle switching can be deterministically achieved regardless of the polarity of the current pulse, in contrast to the quasi-static Slonczewski torque-induced regime as in Fig. 1(b).

### B. Torques dependence

We now extend our study to a large range of the values of $c^{\parallel}/c^{\perp}$ and applied current. Figure 3(a) shows the switching behavior depending on the magnitude of the different torques, with all other parameters being kept identical to that in Fig. 1(d). The pulse is 20 ns long and has infinitely short rising and falling times. The horizontal axis and vertical axis correspond to the strengths of the field-like ($h^{\perp}$) and the Slonczewski ($h^{\parallel}$) torques, respectively. Blue color denotes the regions where magnetization switching is achieved, while red color denotes the regions where the magnetization remains in its initial state. For the low positive values of the torques ratio ($c^{\parallel}/c^{\perp} < \sim 0.01$), the damping of the trajectory arising from the Slonczewski torque remains low. Then many precessions around $\hat{\boldsymbol{\sigma}}$ occur, and switching is non-deterministic as the final position after 20 ns is still not stabilized and cannot be predicted. This corresponds to the mixed blue and red horizontal stripe in Fig. 3(a). However, when both torques are of same sign with a significant Slonczewski torque ( $c^{\parallel}/c^{\perp} > \sim 0.01$ ), the precession number $N$ can be determined as **m** reaches a final, stable position before the falling edge of the pulse, and deterministic switching (or non-switching) is observed.

Depending on $N$, **m** switches via half-precession ( $N = 1/2$ ), switches via more than one precession ( $N = 3/2, 5/2, ...$ ), does not switch after some precessions ( $N = 1, 2, ...$ ), or does not attempt switching at all ( $N = 0$ ). The particular values of the torques $h^{\perp}$ and $h^{\parallel}$ in Fig. 1(d) and Figs. 2(b)-2(f) are marked by the white triangles in Fig. 3(a). When the torque resulting from the combination of $h^{\perp}$ and $h^{\parallel}$ is too strong, the 'up' and 'down' equilibria merge with the equilibrium located at $\hat{\boldsymbol{\sigma}}$, and the stable **m** under currents lies in-plane along $\hat{\boldsymbol{\sigma}}$. Due to the symmetry of the system, the subsequent relaxation is non-deterministic with equal probabilities of relaxation either along $+\hat{\mathbf{z}}$ or $-\hat{\mathbf{z}}$ (considering thermal fluctuations). The white domains in Fig. 3(a) correspond to when **m** lies in-plane under currents. In the absence of Slonczewski torque, the condition for 'up' and 'down' merging is derived as $h^{\perp} \geq 1$.

Among these outcomes, the most interesting case is switching via half-precession. This case corresponds to the blue regions labeled (1/2) in Fig. 3(a). This region extends to a significant range of the diagram, therefore switching via half precession is quite robust with respect to the relative values of $c^{\parallel}$ and $c^{\perp}$. Moreover, for some values of $c^{\parallel}/c^{\perp}$, we observe that $N$ is limited to 0 or $1/2$. The case of $c^{\parallel}/c^{\perp} = 0.25$, for example, is shown by the dashed line in Fig. 3(a). We can further compare this scheme with field induced switching using pulses of undefined length, for which the suitable current intensity is strongly linked to $\alpha$ [37]. Because $\alpha$ is fixed by the material choice in a device, the window of the switching field amplitudes is limited. In our situation, on the contrary, where the damping arises from the Slonczewski-torque, the effective damping is proportional to the current. For greater switching currents, that is, greater initial lags, the effective damping is also stronger and balances the lag so that switching can still be achieved by half-precession.

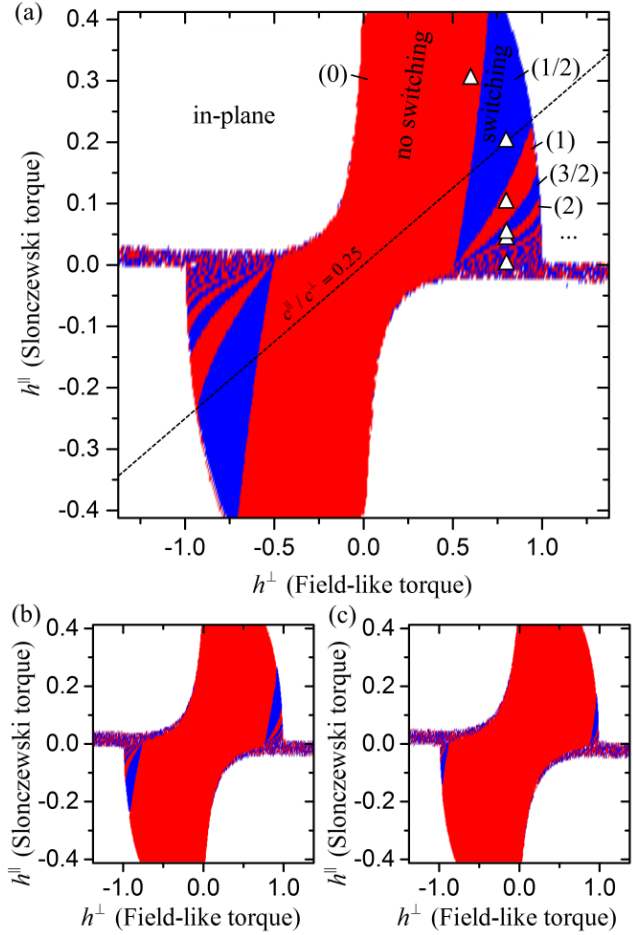

FIG. 3. Final position at the end of the pulse as a function of the strengths of the Slonczewski and field-like torques with $t_{rise} = 0$ (a), $t_{rise} = 300$ ps (b), and $t_{rise} = 1000$ ps (c). Red regions indicate unswitched, blue color represents switched, and white domains denote when **m** lies in-plane. In (a), the dashed line shows the torques that correspond to $c^{\parallel}/c^{\perp} = 0.25$. Triangles indicate the different cases shown in Fig. 1(d) and Fig. 2. Numbers in brackets are the precession number $N$.

Having $N \leq 1/2$ on the whole range of current densities allows a reliable switching behavior regardless of the variations of the amplitude of the current. The required range of $c^{\parallel}/c^{\perp}$ to observe this regime without switching back is found to be 0.23 – 0.55, with $\alpha = 0.01$. Interestingly, among the several material combinations studied so far, the ratio $c^{\parallel}/c^{\perp}$ of Ta/CoFeB/MgO has been reported to be in this range [24,26,27] as well as that of Hf/CoFeB/MgO [28]. It is noteworthy that if the sign of $c^{\parallel}$ and $c^{\perp}$ are not same ($c^{\parallel}/c^{\perp} < 0$), reliable switching is not observed or the magnetization stabilizes in the $xy$-plane, as indicated by the white color in the second and fourth quadrants in Fig. 3(a). Note that the switching characteristics are not noticeably modified for values of $t_{rise} \leq 50$ ps, but are altered for larger $t_{rise}$ values, as can be seen from Figs. 3(b) and 3(c) corresponding to $t_{rise} = 300$ ps and $t_{rise} = 1000$ ps, respectively. We thus need to consider the influence of the rising time on switching.

## C. Influence of rising time

So far we have discussed the field-like torque dominated switching by modeling the current pulse with $t_{rise} = 0$. This is an ideal case, and switching strongly relies on the initial lag of the magnetization. In the case of a finite rising time, the switching behavior can be altered. Figure 4(a) shows the switching trajectories with $h^{\parallel} = 0.2$ and $h^{\perp} = 0.8$, with different values of $t_{rise}$. For the shorter rising times (blue and green lines corresponding to $t_{rise} = 50$ ps and 200 ps, or $t'_{rise} \approx 0.3\pi$ and $1.2\pi$, respectively), the magnetization reverses by half-precession as expected. For $t_{rise} = 50$ ps, the trajectory is not noticeably changed with respect to the case of an infinitely sharp pulse, but the point where **m** crosses the *xy*-plane is displaced towards the saddle point located in the direction of $\hat{\boldsymbol{\sigma}}$ when $t_{rise}$ further increases to $t_{rise} = 200$ ps. For the longer $t_{rise}$ (orange and red lines corresponding to $t_{rise} = 400$ ps and 800 ps, or $t'_{rise} \approx 2.2\pi$ and $4.5\pi$, respectively), **m** evolves significantly and the additional energy is dissipated during the rising edge of the pulse. This reduces the initial amplitude of the trajectory so that **m** does not reach the *xy*-plane and cannot switch. As a consequence, for the same pulse amplitude, the longer rising times lead to less efficient switching or even prevent switching.

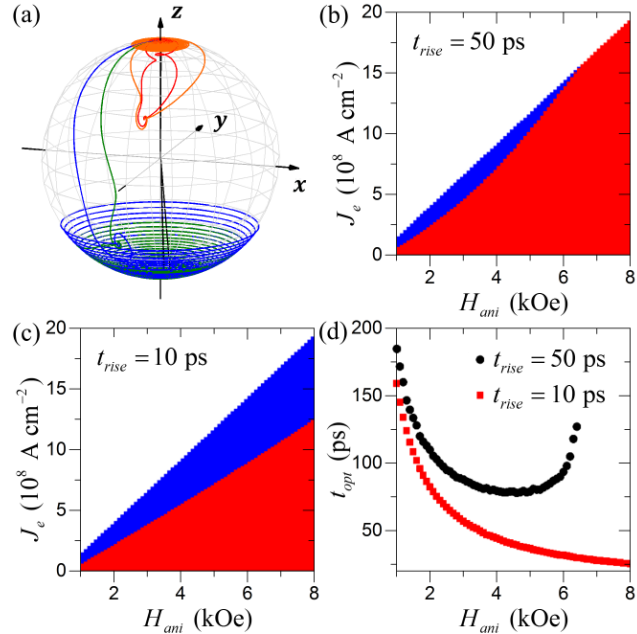

FIG. 4. (a) Switching trajectories for different $t_{rise}$ (from left to right: blue - 50 ps, green - 200 ps, orange - 400 ps, and red - 800 ps) with $h^{\parallel} = 0.2$ and $h^{\perp} = 0.8$. Switching behavior depending on $H_{ani}$ and $J_e$ for $t_{rise} = 50$ ps (b) and $t_{rise} = 10$ ps (c). Red regions indicate unswitched and blue color represents switched regimes, while white color denotes current densities for which **m** lies in-plane during the pulse. (d) Optimal switching time ($t_{opt}$) versus anisotropy for the two $t_{rise}$ values. For (b)-(d), $c^{\parallel} = 0.07$ and $c^{\perp} = 0.28$.

### D. Switching times

We now discuss the achievable switching times. According to the last term of Eq. (4), the frequency of the current-induced precession around $\hat{\boldsymbol{\sigma}}$ is roughly proportional to the field-like torque term, and thus the switching speed is also expected to be proportional to the current density $J_e$. However, we have seen earlier that once the SOT effective field overcomes the anisotropy field, **m** aligns with $\hat{\boldsymbol{\sigma}}$ and switching is subject to a stochastic behavior. It is thus not possible to increase $J_e$ arbitrarily in order to induce faster precession around $\hat{\boldsymbol{\sigma}}$ and reduce the switching time. Therefore, other means are required to reduce the switching time. When varying $H_{ani}$, only the scale factor between $t$ and $t'$ as well as $J_0$ are modified. As a consequence, multiplying both $H_{ani}$ and $J_e$ by a common factor, Eq. (4) and all parameters remain unchanged so that the switching trajectory is identical, with faster switching. For the switching to be exactly reproduced, any rising time of the pulse should also be reduced accordingly, however $t_{rise}$ is actually an experimental and practical limitation that is fixed. Increasing $H_{ani}$ up to $t'_{rise} \approx 2\pi$, the initial lag is altered and switching is no longer achievable as we have seen earlier. Nevertheless, in the range of anisotropies that allow switching for a given $t_{rise}$, one can increase $H_{ani}$ to achieve faster, almost identical switching at a higher $J_e$.

Figures 4(b) and 4(c) compare the range of suitable switching current densities for $H_{ani} = 1-8$ kOe, with $t_{rise} = 50$ ps and 10 ps, respectively. With the torque efficiencies $c^{\parallel} = 0.07$ and $c^{\perp} = 0.28$ ($c^{\parallel}/c^{\perp} = 0.25$) chosen in accord with experimental findings [24,26,27], a current density $J_e = 2.1 \times 10^8$ A cm$^{-2}$ at $H_{ani} = 1$ kOe corresponds to the values $h^{\parallel} = 0.2$ and $h^{\perp} = 0.8$ previously considered in Figs. 1(d) and 4(a). It is experimentally feasible to achieve $t_{rise}$ = 50 ps [40] which would require state of the art electronics and a well calibrated device design, while $t_{rise}$ = 10 ps is close to an ideal case. At $H_{ani}$ = 1 kOe, $t_{rise}$ = 50 ps and 10 ps correspond to $t'_{rise} \approx 0.28\pi$ and $0.06\pi$, respectively, while at $H_{ani}$ = 8 kOe, they correspond to $t'_{rise} \approx 2.25\pi$ and $0.45\pi$, respectively. As in Fig. 3, red corresponds to an unswitched final state, blue indicates proper switching, and white color denotes when the current density is too high for the anisotropy and drives **m** in-plane. It is clear in Fig. 4(b) that the critical current density for switching is proportional to $H_{ani}$ up to $H_{ani}$ = 6.5 kOe, beyond which switching cannot be achieved because $t'_{rise} \approx 2\pi$.

When $t_{rise}$ is shortened from 50 to 10 ps, a clear switching window whose limits are proportional to $H_{ani}$ is obtained even for the higher anisotropies as $t'_{rise} \ll 2\pi$. As Figs. 4(b) and 4(c) are modified by the finite rising time of the pulse, one can also understands the role of $t_{rise}$ in Fig. 3. Increasing $t_{rise}$ decreases the initial lag of **m** and the initial energy shift, so that the energy provided by the field-like torque should be increased and the energy dissipation by the Slonczewski torque should be reduced accordingly, in order to succeed to switch **m**. That causes the different switching domains in Fig. 3(a) to shrink. All the different domains with $N \geq 1/2$ are greatly reduced with $t_{rise}$ = 300 ps ($t'_{rise} \approx 1.7\pi$) as can be seen in Fig. 3(b). They are even almost suppressed for $t_{rise} \geq 1000$ ps ($t'_{rise} \geq 5.5\pi$), and the central part is entirely occupied by the non-switching red domain $N = 0$ [Fig. 3(c)], as expected in the almost quasi-static regime.

As mentioned earlier, increasing the anisotropy constant aims at reducing the switching times. Figure 5(a) shows, for example, the time evolution of **m** with $H_{ani} = 1$ kOe, $J_e = 2.1 \times 10^8$ A cm$^{-2}$, and $t_{rise} = 50$ ps. The switching time $t_{sw}$ is defined as the time difference between the start of the rising edge of the current pulse and the time when **m** crosses the *xy*-plane, as represented in Fig. 5(a). As shown in Fig. 5(b), increasing both $H_{ani}$ and $J_e$ threefold causes $t_{sw}$ to decrease from 170 to 90 ps following a similar trajectory. However, $t_{sw}$ is not shortened by a factor of 3 because of the fixed value of $t_{rise} = 50$ ps.

We can relate the evolution of the window shown in Fig. 4(b) and the evolution of $t_{sw}$. Within the range of current densities that enables switching at a given $H_{ani}$, the value of $t_{sw}$ can slightly vary. For the purpose of comparison between the different anisotropies, we define at each value of $H_{ani}$ the optimal switching time $t_{opt}$ as the value of $t_{sw}$ obtained for the central $J_e$ in the range of properly switching currents. Figure 4(d) shows $t_{opt}$ as a function of $H_{ani}$, for both $t_{rise} = 50$ and 10 ps. The smallest switching time is $t_{opt} \approx 80$ ps with $t_{rise} = 50$ ps, obtained near $H_{ani} = 4$ kOe and $J_e = 8.4 \times 10^8$ A cm$^{-2}$. With $t_{rise} = 10$ ps, a shorter switching time ($t_{opt} \approx 25$ ps and less) can be obtained by increasing $H_{ani}$ to 8 kOe and more. As expected, $t_{opt}$ is initially inversely proportional to $H_{ani}$ as $t'_{rise} \ll 2\pi$, as can be seen from the black curve ($H_{ani} = 1-3$ kOe), or from the red curve in Fig. 4(d). Then $t_{opt}$ with $t_{rise} = 50$ ps reaches a minimum value before it increases with the anisotropy. This trend corresponds to the narrowing of the window of switching currents that can be seen in Fig. 4(b). The switching time $t_{opt}$ increases more and more rapidly as $H_{ani}$ reaches the limit value for which $t'_{rise} \approx 2\pi$, beyond which no switching is observed. For a particular value of $t_{rise}$, $J_e$ and $H_{ani}$ can be tuned for the smallest $t_{opt}$. As long as $t_{rise}$ is much shorter than $t_{opt}$, the switching time evolves as $1/H_{ani}$ while the required $J_e$ is proportional to $H_{ani}$, so that increasing the anisotropy to achieve faster switching does not require more energy to be dissipated in the device.

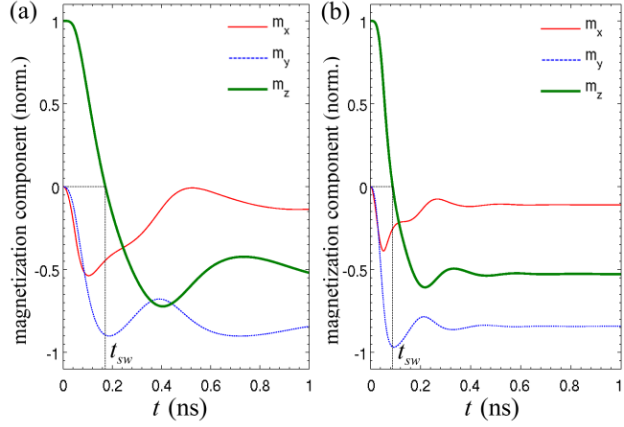

FIG. 5. Time evolution of **m** with $t_{rise} = 50$ ps, $c^{\parallel} = 0.07$, and $c^{\perp} = 0.28$ for (a) $H_{ani} = 1$ kOe and $J_e = 2.1 \times 10^8$ A cm$^{-2}$, and (b) $H_{ani} = 3$ kOe and $J_e = 6.3 \times 10^8$ A cm$^{-2}$.

## IV. MICROMAGNETIC SIMULATIONS

We finally discuss the validity of the macrospin model regarding the size of the device, as the macrospin behavior is an essential assumption in the study presented above. We have performed micromagnetic simulations of SOT switching without assist field. We now include the non-uniform magnetization contribution that can be expressed as

$$E_{n.u.} = A_{ex}\left(\left(\frac{\partial \mathbf{m}}{\partial x}\right)^2 + \left(\frac{\partial \mathbf{m}}{\partial y}\right)^2\right) + D\left(m_z \frac{\partial m_x}{\partial x} - m_x \frac{\partial m_z}{\partial x} + m_z \frac{\partial m_y}{\partial y} - m_y \frac{\partial m_z}{\partial y}\right) \quad (5)$$

where the first term is the exchange energy density ($A_{ex}$ is the exchange constant) and the second term is the DMI energy density ($D$ is the DMI parameter). The total energy density is then given at each point of the ferromagnet by

$$E_{tot} = K_u \left(1 - (\mathbf{m} \cdot \hat{\mathbf{z}})^2\right) - \frac{1}{2}\mu_0 M_S \mathbf{H}_{\mathbf{demag}} \cdot \mathbf{m} + E_{n.u.} \qquad (6)$$

where $K_u$ is the perpendicular uniaxial anisotropy and $\mathbf{H}_{\mathbf{demag}}$ is the magnetostatic demagnetizing field. From Eq. (5) and (6), it appears that the DMI can stabilize a non-uniform magnetization inside the nanomagnet, opposing the exchange coupling. In order to compare with the results of the above macrospin study, the same simulation parameters are used with an exchange stiffness $A_{ex} = 3 \times 10^{-11}$ J m$^{-1}$ [41]. Among various materials, the DMI constant was experimentally determined as $|D| \approx 0.053$ mJ m$^{-2}$ for Ta/CoFe/MgO and $|D| \approx 1.2$ mJ m$^{-2}$ for Pt/CoFe/MgO [42], while $|D| \approx 1.0$ mJ m$^{-2}$ for Pt/CoFeB/MgO [43]. For Hf or W/CoFeB/MgO, different thickness dependent values up to $|D| \approx 0.5$ mJ m$^{-2}$ have been reported [44]. We therefore consider either no DMI ($D = 0$) or an intermediate DMI strength ($D = 0.5$ mJ m$^{-2}$) for this study.

When the DMI is absent ($D = 0$) and for a device size of $50 \times 50 \times 1.6$ nm$^3$ (cell size of $2.5 \times 2.5 \times 0.8$ nm$^3$), we clearly observe a reversal of the magnetization through uniform motion, in particular when no assist field is applied. Increasing $c^{\parallel} / c^{\perp}$ from 0 to 0.5, we find various switching regimes such as non-deterministic regime with many successive switching and switching back, half-precession deterministic switching and no-switching, consistently with the results in Fig. 3(a). We also observe that the higher current densities drive the magnetization in-plane.

We have further studied the switching behavior of larger samples to understand why switching without assist field is not observed in the experiments conducted so far. Most previous studies used large patterned samples ($> 200 \times 200 \text{ nm}^2$), in which the DMI may also play a role [20]. We have simulated switching for a wide range of current densities, in square nanomagnets with the size $w = 50, 100,$ and $200 \text{ nm}$. As can be seen from Eq. (6), the magnetostatic field $\mathbf{H}_{\text{demag}}$ can compete with the anisotropy, reducing the strength of the effective anisotropy field $H_{ani}$. In order to get switching currents comparable with the previous macrospin study for each size, we compensate the increasing demagnetization energy term with $w$ by increasing the uniaxial magnetic anisotropy energy $K_u$ as well, so that the effective anisotropy remains $H_{ani} = 1000$ Oe for all sizes. During the relaxation phase after the pulse, the damping parameter is set to $\alpha = 0.3$ for faster convergence. Note that because of the different demagnetizing fields in the center and in the corners, the switching behavior is no longer ideal. In particular, $c^{\parallel} / c^{\perp} = 0.3$ is used to avoid switching back from the edges in some cases.

Figures 6(a)-6(f) show the mean value of the $m_z$ component at the end of the current pulse (green circles) and after relaxation (red stars) as a function of $J_e$ for various $w$ sizes and DMI parameters. The inset of each figure is a map (top view) of the $m_z$ component in the ferromagnet just before the falling edge of the pulse for a current density of $1.43 \times 10^8$ A cm$^2$, with the corresponding point indicated by an arrow in each figure. Red, white, and blue regions correspond to the $m_z$ values indicated by the color bar in Fig. 6(d). Without DMI [Figs. 6(a)-6(c)] the switching process leads to a uniform state that relaxes to $\mathbf{m} = -\hat{\mathbf{z}}$ after the pulse. As can be inferred from the large white area

in the inset of Fig. 6(c), the magnetostatic energies become more important with increasing device sizes, leading to non-uniform magnetization and non-coherent motion in the larger devices. This alters the switching process as the macrospin behavior is no longer valid, and significantly reduces the window of effective switching currents. With $w = 200$ nm, switching is observed only for high current densities that drive **m** almost in-plane, where thermal fluctuations may disrupt the reliable switching. As a consequence, deterministic switching without assist field is not expected in larger nanomagnets.

Furthermore, the detrimental influence of $w$ on switching is reinforced by the contribution of the DMI. The DMI leads to domain wall formation and non-uniformity of the magnetization within the ferromagnetic layer. In the case of $w = 50$ nm with $D = 0.5$ mJ m$^2$, coherent switching through quasi-uniform motion of **m** [whose state at the end of the pulse is represented in the inset labeled (i) of Fig. 6(d)] can occur only for limited current densities. For higher current densities, as **m** is closer to the $xy$-plane, the DMI overcomes the exchange interaction and leads to a multidomain final state that can be seen from the inset labeled (ii) of Fig. 6(d). In the cases of $w = 100$ nm and $w = 200$ nm [Figs. 6(e) and 6(f), respectively], the DMI disrupts the uniform motion of **m** and no switching is achieved. Therefore, switching in the absence of external assist field is expected only in small devices where the DMI is weak, for example, using Ta for the HM layer. The influence of the DMI may be reduced, when $H_{ani}$ is much higher. However, the higher the anisotropy is, the shorter the rise time $t_{rise}$ has to be. When the effective perpendicular anisotropy is of the order of 4 kOe or more, in most experiments conducted so far (e.g., [19]), $t_{rise}$ should be ~ 50 ps or less. In addition to the detrimental influences of $w$ and $D$, larger rising times in experiments will prevent switching, which may explain why coherent switching driven by the field-like torque has not been experimentally achieved.

## V. SUMMARY

We propose a regime of magnetization switching exploiting the field-like spin-orbit torque, so that switching does not require any assist field. When the relative efficiencies of the torques match a required ratio (0.23-0.55) between the Slonczewski and the field-like torques, magnetization reversal can occur deterministically through a half precession trajectory within tens of picoseconds. This regime is found to be insensitive to the length of the applied pulse, thus allowing reliable switching. When the rise time of the current pulse is short enough, a higher anisotropy ferromagnet allows faster switching for the same pulse energy in the limit of the single-domain behavior. The proposed current-induced SOT switching may be compatible, for example but not limited, with the Ta based trilayer systems. The possibility of achieving reliable current-induced SOT switching without applying external fields paves a way to utilize SOT switching for memory device applications.

This research is supported by the National Research Foundation (NRF), Prime Minister's Office, Singapore, under its Competitive Research Programme (CRP award no. NRFCRP12-2013-01).

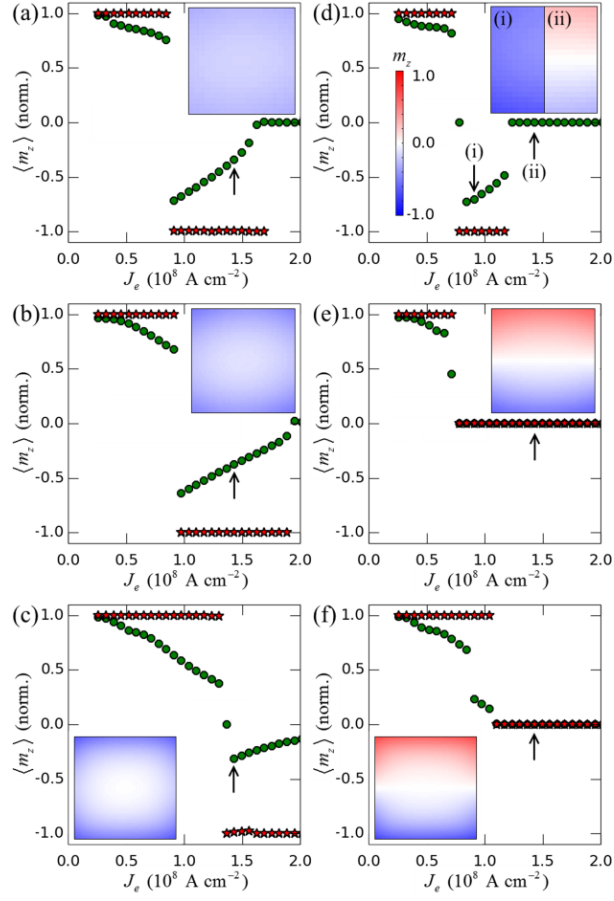

FIG. 6. Mean value of the $m_z$ component of **m** at the end of the current pulse (green circles) and after relaxation (red stars) depending on $J_e$ for (a) $w = 50$ nm and $D = 0$, (b) $w = 100$ nm and $D = 0$, (c) $w = 200$ nm and $D = 0$, (d) $w = 50$ nm and $D = 0.5 \times 10^{-3}$ J m$^{-2}$, (e) $w = 100$ nm and $D = 0.5 \times 10^{-3}$ J m$^{-2}$, and (f) $w = 200$ nm and $D = 0.5 \times 10^{-3}$ J m$^{-2}$. The inset of each figure is a map of the $m_z$ component in the ferromagnet immediately before the falling edge of the pulse for a current density of $1.43 \times 10^8$ A cm$^{-2}$, with the corresponding point indicated by an arrow in each case. Colors correspond to the $m_z$ values indicated by the colorbar in (d). In the particular case of (d), two arrows labeled (i) and (ii) indicate two different situations under current densities of $0.91 \times 10^8$ A cm$^{-2}$ and $1.43 \times 10^8$ A cm$^{-2}$, respectively.